\documentclass[aps,prd,letterpaper,showkeys,showpacs,twocolumn,nofootinbib,preprintnumbers,longbibliography,unsortedaddress]{revtex4-1}

\usepackage[top=2.8cm, bottom=2.8cm, left=2.4cm, right=2.4cm]{geometry}
\usepackage[utf8]{inputenc}  
\usepackage{amsmath,amssymb,lmodern,amsfonts} 
\usepackage{graphicx}
\usepackage{hyperref}
\usepackage{color}
\usepackage{natbib}
\usepackage[caption=false]{subfig}

\newcommand{\overbar}[1]{\mkern 1.5mu\overline{\mkern-1.5mu#1\mkern-1.5mu}\mkern 1.5mu}

\begin{document}

\title{On the coupling of vector fields to the \\ Gauss-Bonnet invariant}

\author{Juan C. Bueno-S\'anchez}
\email{juancarlos.bueno@upm.es}
\affiliation{Departamento de Ingenier\'ia El\'ectrica, Electr\'onica, Autom\'atica y F\'isica Aplicada, \\ Universidad Polit\'ecnica de Madrid, Madrid, Espa\~na}

\author{J. Bayron Orjuela-Quintana}
\email{john.orjuela@correounivalle.edu.co}
\affiliation{Departamento  de  F\'isica,  Universidad  del Valle, \\ Ciudad  Universitaria Mel\'endez,  Santiago de Cali  760032,  Colombia}

\author{C\'esar A. Valenzuela-Toledo}
\email{cesar.valenzuela@correounivalle.edu.co}
\affiliation{Departamento  de  F\'isica,  Universidad  del Valle, \\ Ciudad  Universitaria Mel\'endez,  Santiago de Cali  760032,  Colombia}

\begin{abstract}

Inflationary models including vector fields have attracted a great deal of attention over the past decade. Such an interest owes to the fact that they might contribute to, or even be fully responsible for, the curvature perturbation imprinted in the cosmic microwave background. However, the necessary breaking of the vector field’s conformal invariance during inflation is not without problems. In recent years, it has been realized that a number of instabilities endangering the consistency of the theory arise when the conformal invariance is broken by means of a non-minimal coupling to gravity.  In this paper, we consider a massive vector field non-minimally coupled to gravity through the Gauss-Bonnet invariant, and investigate whether the vector can play the role of a curvaton while evading the emergence of instabilities and preserves the large-scale isotropy.
\end{abstract}



\maketitle

\section{Introduction} \label{sec:01}

Thanks to a wealth of high precision cosmological observations, specially those obtained by the WMAP \cite{Bennett12,Hinshaw12,Komatsu10,Spergel:2006hy,Peiris:2003ff} and Planck missions \cite{Aghanim:2018eyx,Akrami:2018odb,Akrami:2019izv}, cosmological inflation is widely recognized as the simplest paradigm to generate the observed adiabatic, nearly scale-invariant, gaussian spectrum of superhorizon fluctuations imprinted in the cosmic microwave background (CMB). In particular, single-field models, in which the inflationary expansion is driven by a scalar field minimally coupled to gravity, are clearly favored by data. Despite their excellent agreement with the available data, indications exist suggesting that single-field models might need to be extended. The most notorious among these indications are the possible presence of the so-called CMB anomalies (see for instance \cite{Perivolaropoulos:2014lua} for an overview of some of them), firstly reported by WMAP  \cite{WMAP7Anom} and later by Planck \cite{Akrami:2019bkn}. However, since the statistical significance of these effects is small, the existence of these anomalies has been debated in the literature and they are yet to be confirmed \cite{Akrami:2019bkn, Schwarz:2015cma}.

Although scalar fields have played a dominant role in inflationary cosmology, over the last decade it has been realized that vector fields may also have an important function provided their conformal symmetry is broken (see for instance \cite{Soda12,Dima10a,Maleknejad12,DeFelice:2016yws,Heisenberg:2016wtr,Heisenberg:2018mxx,Heisenberg:2018vsk} and references therein). This breaking, which can be brought about by the introduction of a mass term, for example, allows the vector field to obtain a superhorizon spectrum of perturbations during inflation. In turn, this opens up the possibility that the vector field becomes a curvaton and contributes to the curvature perturbation \cite{Lyth:2001nq, Dimopoulos06, Dimopoulos08a, Dimopoulos08b, Dimopoulos09a, Dimopoulos09vu, Dimopoulos11, navarro13a, Karciauskas10as, Dimopoulos10xq}, for which the vector field must come to dominate (or nearly dominate) the energy density at a later epoch. However, the risk when considering the influence of vector fields in the cosmological dynamics is that, since they signal a preferred direction in space, they may result in an anisotropic expansion in excess of the current observational bounds. To quantify the level of statistical anisotropy\footnote{We distinguish between background anisotropy from statistical anisotropy, given that the latter is perturbative in origin.} it is usual to parametrize the spectrum of the curvature perturbation as \cite{Ackerman07}
\begin{equation} \label{Ackerman P}
{\cal P}_\zeta(\mbox{\boldmath $k$})={\cal P}_\zeta^{\rm iso}(k)
[1+g(k)(\mbox{\boldmath $d\cdot\hat k$})^2]\,,
\end{equation}
where ${\cal P}_\zeta^{\rm iso}(k)$ denotes the isotropic part of the power spectrum, $g(k)$ is the so-called anisotropy parameter, which quantifies the statistical anisotropy in the spectrum ${\cal P}_\zeta$, \mbox{\boldmath $d$} is the unit vector signaling the preferred direction, and $\mbox{\boldmath $\hat k$}\equiv\mbox{\boldmath $k$}/k$ is the unit vector along the wavevector \mbox{\boldmath $k$} with modulus $k$. Observations from the Planck satellite suggests that $g$ can be at most 0.97 \cite{Ade:2015hxq},  which represents a very tight constraint on the contribution of vector fields to the power spectrum of the CMB. Nevertheless, if the isotropy of the expansion is approximately preserved, vector fields could even be responsible for inflation \cite{Golovnev08,Golovnev09a,Golovnev09b,Golovnev:2011yc,Emami:2016ldl}. The requirement in this case is to have either a large number (typically in the hundredths) of randomly oriented vector fields so that they collectively generate a nearly isotropic expansion, or consider three mutually orthogonal vector fields with equal vev \cite{Golovnev08}. It is also possible to retain an isotropic inflationary expansion by considering the dynamics of gauge vector fields. This is the case of gaugeflation \cite{Maleknejad12,Maleknejad11a,Maleknejad11b,Nieto:2016gnp,Rodriguez:2017wkg,Adshead:2017hnc}, in which a nonabelian gauge field minimally coupled to gravity plays the role of the inflaton. In this proposal, an SU(2) gauge field is considered to form a triad of mutually orthogonal vectors, which in turn allows the gauge field to drive inflation without giving rise to an anisotropic expansion.

An interesting manner to break the conformal invariance is by considering a non-minimally coupled vector field, thus resulting in a modification of gravity \cite{Dimopoulos08a,Golovnev08,Golovnev09a,Golovnev09b}. Unfortunately, the non-minimal coupling to gravity is known to be problematic due to the emergence of instabilities \cite{Himmetoglu08a,Himmetoglu08b,Himmetoglu09a}. Although the existence of these problems represents a serious drawback for the consistency of the theory, the very nature of the instabilities has been called into question, and a number of scenarios have been envisaged to evade them \cite{Karciauskas10as}. In this paper, we examine a cosmological vector field non-minimally coupled to gravity through the Gauss-Bonnet invariant. Although couplings between the Gauss-Bonnet invariant and scalar fields have been explored in the context of inflationary cosmology \cite{Tsujikawa:2004dm,Carter05fu,Satoh08ck,Guo09uk,Sadeghi09pu,Guo10jr,Satoh10ep,Jiang13gza,Nozari13wua,Koh:2014bka,Okada:2014eva,Kanti:2015pda,vandeBruck:2015gjd,vandeBruck:2016xvt,Koh:2016abf,Mathew:2016anx,vandeBruck:2017voa,Fomin:2017qta,Yi:2018gse,Granda:2019jqy,Granda:2019wyi,Jimenez:2019gmr,Kleidis:2019ywv,Fomin:2020hfh,Pozdeeva:2020shl,Rashidi:2020wwg},  the coupling with a massive vector field has not been sufficiently explored in the literature\footnote{Non-minimal couplings of the electromagnetic field to gravity, in particular to the Gauss-Bonnet invariant, have been considered as a mechanism to generate large-scale magnetic fields during inflation \cite{Sadeghi09pu, Bamba:2008ja}.} \cite{Oliveros:2016myr}. Arguably, this is due to the very presence of instabilities in relatively simple settings, as in the case of a non-minimal coupling to the Ricci scalar, which then invites to exercise caution when considering more complicated non-minimal couplings. Nevertheless, the reason for us to invoke such a coupling owes to the peculiar behavior of the Gauss-Bonnet invariant. Indeed, a very crucial feature is that it changes its sign when passing from inflation to a matter or radiation dominated phase. Consequently, a mass term for the vector field, coming from this coupling, features the same change of sign towards the end of inflation. In the vector curvaton scenario  \cite{Dimopoulos06}, a negative mass-squared is required for the vector field to generate a nearly flat power spectrum, while a positive mass-squared is required for the vector field engages into quick oscillations after inflation, in order to avoid the generation of large background anisotropies if the vector field dominates the Universe. However, a clear mechanism producing this change of sign is not provided, and thus it has to be assumed \cite{Dimopoulos06}. In this regard, the goal of this paper is to investigate whether a vector field coupled to this topological term can contribute significantly to the total energy density after inflation, thus being able to play the role of a curvaton. Two key assumptions are made in order to keep the isotropy in the expansion: $i)$ the vector field is subdominant during inflation \cite{Dimopoulos06} and $ii)$ after inflation the vector field conduct itself as a pressureless mater \cite{Dimopoulos06}.  These assumptions allow us to safely use an isotropic and homogeneous spacetime, i.e. the Friedmann-Lema\^itre-Robertson-Walker (FLRW) metric. 

The paper is organized as follows. In Sec. \ref{stability}, we clarify our position with respect to the instability of the theory. The Lagrangian density for a vector field coupled to the Gauss-Bonnet invariant is introduced in Sec. \ref{model}. In Sec. \ref{espectrum}, we compute the perturbation spectrum and the anisotropy parameter $g$. Section \ref{EVF} is devoted to study the dynamics of the vector field during and after inflation. In Sec. \ref{energy}, we study the evolutiton of the energy density and show that the condensate of an oscillating heavy vector field behaves like a pressureless fluid. It means that the vector does not generate large background anisotropies, fulfilling the requirements to be a suitable curvaton field. Finally, we present our conclusions in Sec. \ref{sec-e}.

\section{On the instability of non-minimally coupled vector fields} \label{stability}

An issue of fundamental importance concerning theories of massive vector fields non-minimally coupled to gravity is their instability \cite{ Karciauskas10as, Himmetoglu08a, Himmetoglu08b, Himmetoglu09a}, which originates from the longitudinal mode of the vector field. One of the known instabilities is perturbative in origin and arises when the effective mass squared of the vector field changes its sign from negative to positive \cite{Himmetoglu08a, Himmetoglu08b, Himmetoglu09a}. In the scenario studied in Refs. \cite{Dimopoulos08b,Karciauskas10as}, it was shown in Ref. \cite{Dimopoulos08b} that the instability is under control during inflation. However, the instability arises at some later epoch, when the field's effective mass squared crosses zero. In spite of this difficulty, the authors in \cite{Karciauskas10as} go on to argue that even if such instability exists, it still might be possible to avoid it if the bare mass of the vector field stems from the coupling to another field, which then would allow either a curvaton or an inhomogeneous reheating mechanism. To the best of our knowledge, the debate on this issue is not yet settled, and hence our attitude towards it will be the same as in Ref. \cite{Karciauskas10as}, thus simply ignoring the instability or assuming that, if present, it can be circumvented by some mechanism. Although this attitude conveniently dispenses with the problem, it is also fair to say that such instability arises when the longitudinal mode becomes unphysical, and hence it is reasonable to suspect that the associated singularity might share the same unphysical nature.

Apart from the above, yet another problem plagues this kind of vector field models, the so-called \emph{ghost instability}. It originates because, during inflation, the kinetic energy density for the longitudinal modes of the vector field have the \emph{wrong} sign, which might entail the copious production of vector field quanta up to the point of ruining inflation. Regarding this instability, the authors in Ref.~ \cite{Karciauskas10as} argue that as long as the negative energy contributed by ghost states does not exceed the energy density driving inflation, these are in principle not problematic for the stability of the theory. In the following, we implicitly assume that this is indeed the case.

\section{Vector field coupled to the Gauss-Bonnet invariant} \label{model}

We consider a massive vector field coupled to the Gauss-Bonnet invariant and evolving in an inflationary background, which we take to be quasi-de Sitter and driven by an unspecified matter source. The action of the system is
\begin{equation} \label{Lagrangian}
\mathcal{L} \equiv \mathcal{L}_\text{EH} + \mathcal{L}_\text{inf} + \mathcal{L}_A + \mathcal{L}_\mathcal{G} \,,
\end{equation}
with\footnote{Greek indices run from 0 to 3 and denote spacetime coordinates. Latin indices run from 1 to 3 and denote spatial components.}
\begin{equation*}
\mathcal{L}_\text{EH} \equiv - \frac{m_\text{P}^2}{2} R \,,\ \mathcal{L}_A \equiv - \frac14 F_{\mu\nu} F^{\mu\nu} + \frac12 \tilde m^2  \, A_\mu A^\mu \,,
\end{equation*}
\begin{equation}
\mathcal{L}_\mathcal{G} \equiv \frac12 \alpha \, {\cal G} \, A_\mu A^\mu \,,
\end{equation}
where, $m_\text{P}$ is the reduced Planck mass, $R$ is the Ricci scalar, $F_{\mu\nu} \equiv \nabla_\mu A_\nu-\nabla_\nu A_\mu$ is the strength tensor associated to the vector field $A_{\mu}$ with bare mass $\tilde{m}$, \mbox{${\cal G} \equiv R^2-4R_{\alpha\beta}R^{\alpha\beta}+R_{\alpha\beta\gamma\delta}R^{\alpha\beta\gamma\delta}$} is the Gauss-Bonnet topological invariant with coupling strength\footnote{ The dimensions of this coupling are $[ \alpha ] = m_{\text{P}}^{-2}$.} $\alpha$, and $R_{\mu\nu}$, $R_{\mu\nu\rho\sigma}$ are the Ricci tensor and the Riemann tensor, respectively. $\mathcal{L}_\text{inf}$ is the Lagrangian density for the energy content responsible for the inflationary period. The effective mass squared of the vector field is defined as
\begin{equation} \label{eq2}
m^2 \equiv \tilde m^2 + \alpha \, {\cal G} \,.
\end{equation}
In the case of the FLRW metric \mbox{$\text{d} s^2 = \text{d} t^2 - a^2 (t) \text{d} \boldsymbol{x}^2 $}, where $a(t)$ is the scale factor and $\boldsymbol{x}$ are the Cartesian spatial coordinates, the Gauss-Bonnet invariant reads\footnote{The overdot denotes a derivative with respect to the cosmic time $t$.}
\begin{equation}
{\cal G} = 24 H^2 (\dot H + H^2) \,,
\end{equation}
where $H = \dot{a} / a$ is the Hubble parameter.

\section{Perturbation spectrum} \label{espectrum}

Having clarified our position with respect to the instability of the theory, we investigate the conditions for which the vector field obtains a nearly scale-invariant spectrum of superhorizon perturbations. 
The equation of motion for the vector field, which is obtained by varying the action of the Lagrangian in Eq. (\ref{Lagrangian}) with respect to $A_\nu$, is
\begin{equation} \label{EoM vector}
\nabla_\mu F^{\mu\nu} + m^2 A^\nu = 0 \,.
\end{equation}
Assuming that inflation homogenizes the vector field; i.e. $\partial_i A_\mu = 0$, it is easy to show that its temporal component, $A_t$, must be zero, while the spatial components $A^i$ obey\footnote{See Appendix \ref{Appendix} for the details of the calculations.}
\begin{equation} \label{EoM A}
\ddot{A}^i + H \dot{A}^i + m^2 A^i = 0 \,.
\end{equation}
Now, we perturb the vector field in the following way:
\begin{equation} \label{Perturbative expansion}
A_i (t, \boldsymbol{x}) \equiv A_i (t) + \delta A_i (t, \boldsymbol{x}) \,,\, A_t (t, \boldsymbol{x}) = \delta A_t (t, \boldsymbol{x}) \,,
\end{equation}
and write the equations of motion for transverse  ($\delta \mathcal{A}^i_{\ \perp}$) and  longitudinal ($\delta \mathcal{A}^i_{\ \parallel}$) modes  as follows (see Appendix \ref{Appendix}),
\begin{equation} \label{Transverse A}
\left[ \partial_t^2 + H \partial_t + m^2 + \left( \frac{k}{a} \right)^2 \right] \delta \mathcal{A}^i_{\perp} = 0 \,,
\end{equation}
\begin{equation} \label{Longitudinal A}
\left[ \partial_t^2 + \left( 1 + \frac{2 k^2}{k^2 + (a m)^2} \right)H \partial_t + m^2 + \left( \frac{k}{a} \right)^2 \right] \delta \mathcal{A}^i_{\parallel} = 0 \,,
\end{equation}
where $\delta \mathcal{A}^i$ are the Fourier modes of $\delta A^i$. In the above equations, we used the fact that $m^2$ is a constant during inflation since $\dot{H}\simeq 0$ and therefore $\mathcal{G} \approx 24 H^4$.

In order to quantize the field, we introduce creation/annihilation $\hat a^\dag / \hat a$ operators for each polarization mode
\begin{eqnarray}
\delta\mbox{$\mathcal{A}^j$} (t, \mbox{\boldmath$x$})&\equiv&
\int\frac{\text{d}^3k}{(2\pi)^3} \sum_\lambda \left[
\mbox{$e^j$}_\lambda(\hat{\mbox{\boldmath $k$}})
\hat a_\lambda(\mbox{\boldmath $k$})
z_\lambda(t, k)e^{-i\mbox{\scriptsize\boldmath$k$}\cdot \mbox{\scriptsize
\boldmath$x$}}\right.\nonumber\\
&&\left.+\mbox{$e$}^{j \, *}_{\ \lambda}(\hat{\mbox{\boldmath $k$}})
\hat a^\dag_\lambda(\mbox{\boldmath $k$})
z^*_\lambda(t,k)e^{i\mbox{\scriptsize\boldmath$k$}\cdot \mbox{\scriptsize
\boldmath$x$}}
\right]\label{eq4}\,,
\end{eqnarray}
where $\lambda = \text{L}, \text{R}$ labels the left and right transverse polarizations and $\lambda = ||$ the longitudinal polarization. Choosing the $\boldsymbol{k}$-direction along the $z$-axis, the polarization vectors $\boldsymbol{e}_\lambda$ can be written as
\begin{equation}
\boldsymbol{e}_\text{L} = \frac{1}{\sqrt{2}}(1, i, 0) \,,\ \boldsymbol{e}_\text{R} = \frac{1}{\sqrt{2}}(1, -i, 0) \,,\ \boldsymbol{e}_{||} =(0,0,1) \,,
\end{equation}
while the commutation rules are
\begin{equation}
\left[\hat a_\lambda(\mbox{\boldmath $k$})\,,\,\hat a^\dag_{\lambda'}(\mbox{\boldmath $k$}')\right]=(2\pi)^3\delta(\mbox{\boldmath $k$}-\mbox{\boldmath $k$}')\delta_{\lambda\lambda'}.
\end{equation}
With all the above, the power spectrum for the $\lambda$-polarized modes $z_\lambda$ is defined by
\begin{equation} \label{Power spectrum}
{\cal P}^z_\lambda(k) = \lim_{k / aH \to 0} \frac{k^3 || z_\lambda ||^2}{2\pi^2}\,.
\end{equation}
In the following subsections we study each polarization individually.

\subsection{Transverse modes} \label{Transverse Modes}

Defining the physical transverse modes as $b_{\text{ L, R}} \equiv z_{\text{ L, R}} / a$ and using the conformal time $\text{d} \eta \equiv \text{d} t / a$, the evolution equation (\ref{Transverse A}) is rewritten as
\begin{equation} \label{eq76}
\left[ \partial_\eta^2 + 2 \mathcal{H} \partial_\eta + 2 \mathcal{H}^2 + (a m)^2 + k^2 \right] b_{\text{ L, R}} = 0 \,,
\end{equation}
with $\mathcal{H} \equiv a H$ being the Hubble parameter in conformal time. This equation can be recast in the form of a Bessel equation whose solution is given in terms of the Hankel functions $H_\nu$ as
\begin{equation} \label{Solutionb}
b_\text{ L, R}(\eta, k) = C_{k} \, a^{-1} \sqrt{-\eta} \, H_\nu (- k \eta) \,,
\end{equation}
where
\begin{equation} \label{eq5}
\nu^2\equiv\frac1 4-\frac{\tilde m^2+\alpha{\cal G}}{H^2} =\frac1 4-\frac{m^2}{H^2}\,.
\end{equation}
The constant $C_{k}$ can be found by matching Eq. (\ref{Solutionb}) with the Bunch-Davies vacuum at early times (when $ - k\eta \to \infty$), obtaining
\begin{equation}
C_{k} = \frac{\sqrt{\pi}}{2} \, \, \Rightarrow  \, \, b_{\text{ L, R}} (\eta, k) \approx \frac{1}{a \sqrt{2 k}} e^{-i k \eta} \,.
\end{equation}
As expected, the modes behave as those for an oscillator.
Now, we are interested in the late time behaviour (when $- k \eta \to 0$) of the physical modes. In this regime, the dominant contribution of the solution in Eq. (\ref{Solutionb}) is 
\begin{equation} \label{eq71}
b_\text{ L ,R} (k) \propto k^{- \nu}\,.
\end{equation}
Replacing the later expression in the power spectrum defined in Eq. (\ref{Power spectrum}), we obtain 
\begin{equation}
{\cal P}_{\text{ L, R}} \equiv \frac{\mathcal{P}^z_{\text{ L, R}}}{a^2} \propto k^{3 - 2\nu} \,,
\end{equation}
which corresponds to the scale dependance of the power spectrum of the physical vector field perturbations. The spectral index is written as
\begin{equation} \label{eq8}
n_{\text{ L, R}} - 1 \equiv \frac{d \, \text{ln} {\cal P}_{\text{ L, R}}}{d \, \text{ln} k} = 3 - 2 \nu \,.
\end{equation}
From Eqs. (\ref{eq5}) and (\ref{eq8}) it is clear that the physical vector field can attain a nearly flat power spectrum if and only if $m^2 \approx -2 H^2$. Consequently, the coupling $\alpha$ must satisfy $\tilde{m}^2 + 24 \alpha H^4 \approx - 2 H^2$. Moreover, if $\tilde{m} \ll H$, the condition for scale-invariance becomes
\begin{equation} \label{Condition alpha}
\alpha H^2 \approx - \frac{1}{12} \,.
\end{equation}
This result must be compared to the one in Refs. \cite{Dimopoulos08a, Himmetoglu09a} where, instead of coupling the vector field to the Gauss-Bonnet invariant, the authors couple the vector field to the Ricci scalar, i.e. $\alpha R A_\mu A^\mu$. In Refs. \cite{Dimopoulos08a, Himmetoglu09a}, it was shown that if the coupling constant is exactly $1/6$ the power spectrum is perfectly flat, besides, the spectrum of each transverse mode is precisely the same as that for a scalar field. Finally, it is important to mention that the requirement $m^2 \approx - 2 H^2$ is precisely one of the possibilities discussed in \cite{Karciauskas10as} to avoid the instabilities mentioned in Sec. \ref{stability}.

\subsection{Longitudinal modes}

Equation (\ref{Longitudinal A}) gives the evolution for the longitudinal modes $z_\parallel$, which, in terms of the conformal time, can be written as
\begin{equation}
\left[ \partial_\eta^2 + \frac{2 \mathcal{H} k^2 }{k^2 + (a m)^2} \partial_\eta + (a m)^2 + k^2 \right] z_{\parallel} = 0 \,.
\end{equation}
Using the conditions $\tilde{m} = 0$ and $m^2 = - 2 H^2$ for a scale-invariant transverse power spectrum, and taking into account that the vacuum boundary condition is modified by 
\begin{equation}
\lim_{- k \eta \rightarrow \infty} z_\parallel = \gamma \frac{1}{\sqrt{2 k}} e^{- i k \eta},
\end{equation}
with $\gamma = \sqrt{(k/ a m)^2 + 1}$ the Lorentz boost factor which takes us from the frame with $\boldsymbol{k} = 0$ to the frame of momentum $\boldsymbol{k} \neq 0$, the solution of the above equation  is \cite{Dimopoulos08b, Dimopoulos09a, Dimopoulos09vu, Dimopoulos11}
\begin{equation}
z_\parallel = \frac{\sqrt{-\eta}}{2} \left[ - k \eta + \frac{2}{k \eta} + 2 i \right] \frac{e^{- i k \eta}}{\sqrt{- k \eta}} \,,
\end{equation}
which at late times ($-k \eta \rightarrow 0$) behaves as $\sqrt{- \eta} (- k \eta)^{ - 3/2}$. Replacing the latter in the power spectrum in Eq. (\ref{Power spectrum}) we get
\begin{equation}
\mathcal{P}^z_\parallel \approx 2 a^2 \left( \frac{H}{2 \pi} \right)^2 \,,
\end{equation}
where we used the approximation $H^2 \approx (a \eta )^{-2}$ which is valid during inflation for a quasi-de Sitter background. As in the transverse case, the physical longitudinal power spectrum $\mathcal{P}_\parallel$ can be obtained by defining the physical longitudinal modes as $b_\parallel \equiv z_\parallel / a$.

\subsection{Statistical anisotropy}

It is known that vector fields introduce inherently a preferred direction and therefore they can introduce large statistical anisotropies in the perturbation spectrum. If this is the case, the model will be ruled out because it is in disagreement with the observational results. For this reason, by using the $\delta N$ formalism \cite{Starobinsky86, Sasaki95, Lyth04dn, Lyth05}, in this section we compute the amount of statistical anisotropy in the spectrum, which is quantified in the parameter $g$ defined in Eq. (\ref{Ackerman P}). We will show that $g$ can be small enough to satisfy the observational bounds \cite{Ade:2015hxq}. 

According to the $\delta N$ formalism, the curvature perturbation $\zeta$ is the difference of the number of $e$-folds $N$ between uniform density and spatially flat slices of spacetime: $\zeta \equiv \delta N$. We will assume that $N$ is function of the scalar field $\phi$ and the vector field: $N = N \left( \phi, A^\mu \right)$. Then, the curvature perturbation can be written as \cite{Dimopoulos11}
\begin{align}
\zeta &= N_\phi \delta \phi + N^i_A \delta A_i + \frac{1}{2} N_{\phi\phi} (\delta \phi)^2 \nonumber \\
 &+ \frac{1}{2} N^i_{\phi A} \delta \phi \delta A_i + \frac{1}{2} N^{i j}_{AA} \delta A_i \delta A_j + \dotsi, 
\end{align}
where $N_\phi \equiv \frac{\partial N}{\partial \phi}$, $N^i_A \equiv \frac{\partial N}{\partial A_i}$, $N_{\phi\phi} \equiv \frac{\partial^2 N}{\partial \phi^2}$, $N^i_{\phi A} \equiv \frac{\partial^2 N}{\partial \phi A_i}$, $N^{i j}_{A A} \equiv \frac{\partial^2  N}{\partial A_i A_j}$, $\delta \phi$ and $\delta A_i$ are the perturbations of the scalar and vector field, respectively. From this result, the power spectrum of the curvature perturbation reads
\begin{equation} \label{spectrum 1}
\mathcal{P}_\zeta (\boldsymbol{k}) = N^2_\phi \mathcal{P}_\phi + N^2_A \left[ \mathcal{P}_+ + \left( \mathcal{P}_\parallel - \mathcal{P}_+ \right) ( \boldsymbol{\hat{d}} \cdot \boldsymbol{k} )^2 \right],
\end{equation}
where $\mathcal{P}_\phi$ denotes the power spectrum of the scalar field, we have defined the even and odd polarizations for the transverse spectra as $2 \mathcal{P}_{\pm} \equiv \mathcal{P}_\text{L} \pm \mathcal{P}_\text{R}$, respectively. We have taken into account that our theory is parity conserving, i.e. $\mathcal{P}_\text{R} = \mathcal{P}_\text{L}$, and thus  $\mathcal{P}_+ = \mathcal{P}_\text{R}$ and $\mathcal{P_{-}} = 0$. In the latter equation, we have  also defined $N_A^2 \equiv || \boldsymbol{N}_A ||^2 \equiv N_A^i N_{A i}$ and $\boldsymbol{\hat{d}} \equiv \boldsymbol{N}_A / N_A$, which defines the preferred direction signaled by the vector field. By comparing the above equation with Eq. \eqref{Ackerman P}, we identify the isotropic part of the spectrum as
\begin{equation}
\mathcal{P}_k^{\text{iso}} (k) = N^2_\phi \mathcal{P}_\phi (k) + N^2_A \mathcal{P}_+ (k),
\end{equation}
and hence the anisotropy parameter $g$ can be written as
\begin{equation}
g = \beta \frac{\mathcal{P}_\parallel - \mathcal{P}_+}{\mathcal{P}_\phi + \beta \mathcal{P}_+}, \quad \beta \equiv \frac{N_A^2}{N_\phi^2} \,,
\end{equation}
where $\beta$ quantifies the relative contribution of the vector field over the scalar field to the modulation of $N$. Now, since the power spectra of the transverse solutions are nearly flat, they are given by $\mathcal{P}_{\text{ L, R}} \approx (H / 2\pi)^2$, and assuming that the potential of the scalar field is sufficiently flat during inflation, such that the power spectrum of the scalar field is also nearly flat at horizon exit \cite{Lyth:2001nq}, i.e. $\mathcal{P}_\phi \approx (H / 2 \pi)^2$, we get
\begin{equation}
g \approx \frac{\beta}{1 + \beta} \approx \beta,
\end{equation}
where we took into account that $N$ is primarily modulated by the scalar field, since the vector field is subdominant during inflation, implying $\beta \ll 1$, then $g\approx \beta \ll 1$. This result agrees with the bounds given by Planck which suggest that $g$ can be at most 0.97 \cite{Ade:2015hxq}.

\section{Evolution of the vector field} \label{EVF}

In this section, we follow the evolution of the homogeneous vector field during and after inflation in order to determine if the vector field is able to play the role of a curvaton \cite{Lyth:2001nq, Dimopoulos06}.

Defining the physical components of the vector $A^i $  as $B^i \equiv A^i / a$ and supposing, by simplicity, that $B_\mu = (0, 0, 0, B)$, the equation of motion (\ref{EoM A}) is rewritten as
\begin{equation} \label{EOM B}
\ddot{B} + 3 H \dot{B} + \left( \dot{H} + 2 H^2 + m^2 \right) B = 0 \,.
\end{equation}
In the following, we solve this equation during and after inflation.

\subsection{Evolution during inflation}

As shown in section \ref{Transverse Modes}, the nearly scale-invariant spectrum of vector perturbations is obtained if the effective mass of the vector field  is $m^2 \approx - 2 H^2$, which remains constant during inflation because $H\simeq {\rm constant}$. Therefore, the solution of Eq. (\ref{EOM B}) during inflation is
\begin{equation} \label{Sol Inf}
B(t) = B_0 + B_1 a^{-3} \,,
\end{equation}
where $B_0$ and $B_1$ are integration constants. The decaying mode in the solution in Eq. (\ref{Sol Inf}) is quickly diluted by inflation, thus the field is nearly constant given that $B \approx B_0$. This means that the vector field remains frozen
and therefore it is not ``erased'' in the inflationary phase.

\subsection{Evolution after inflation} \label{evoaf}

The post-inflationary evolution of the vector field is also described by Eq. (\ref{EOM B}), but now $H$ is time depending. Assuming  that $H$ evolves as
\begin{equation} \label{Hubble}
H(t) = \frac{2 \, t^{-1}}{3(1 + w)} \,,
\end{equation}
where $w$ is the equation of state parameter of the dominant fluid after inflation and that $\alpha H^2 = \gamma$, where $\gamma$ is a constant (not necessarily the same required for a flat power spectrum),\footnote{We assume this condition for simplicity, and having in mind the condition in Eq. \eqref{Condition alpha} which is valid during inflation.} Eq. (\ref{EOM B}) can be recast in the following form
\begin{equation}
t^2 \ddot{B} + \frac{2 \, t}{1 + w} \dot{B} + \left[ (\tilde{m} t)^2 - \tau^2 \right] B = 0 \,,
\end{equation}
where
\begin{equation}
\tau^2 \equiv \frac{2 (1 + 3w)}{3 (1 + w)^2} \left( 8 \gamma - \frac{1 - 3w}{3 (1 + 3w)} \right) \,.
\end{equation}
The general solution of the above equation is
\begin{equation} \label{Bessel Sol}
B (t) = t^u \left[ c_1 J_v (\tilde{m}\, t) + c_2 Y_v (\tilde{m}\, t) \right] \,,
\end{equation}
where
\begin{equation}
u \equiv \frac{w - 1}{2 (w + 1)} \, , \ v \equiv \frac{\sqrt{1 + 3w}}{6 (1 + w)} \sqrt{1 + 3w + 192 \gamma} \,,
\end{equation}
$c_1$ and $c_2$ being constants of integration and $J_v$ and $Y_v$ being the Bessel functions of first and second kind, respectively. This solution should be contrasted with the solution for the equation of motion of a vector field non-minimally coupled to the Ricci scalar. In Ref. \cite{Dimopoulos08a}, it was shown that, since this coupling is negligible after inflation\footnote{This is true if a radiation dominated period follows after the inflationary phase, since $R \approx 0$ for an equation of state $w \approx 1/3$.}, the vector field behaves as a massive minimally-coupled abelian vector. In contrast, in our model, the Gauss-Bonnet coupling contributes to the effective mass after inflation as well. This dependence is very important because, since the Gauss-Bonnet invariant changes its sign when passing from inflation to a matter or radiation dominated phase, naturally entails the change of sign of the vector mass. In Refs. \cite{Dimopoulos06, Dimopoulos08a, Dimopoulos09a}, it was shown that a vector field with positive mass after inflation can engage into quick oscillations, avoiding the generation of large-scale anisotropies. However, this sign change has to be assumed since a mechanism provoking this feature is not presented.

Now, we consider two possible approximations of the solution (\ref{Bessel Sol}) regarding the dependence of the Bessel functions with respect to the bare mass $\tilde{m}$ of the vector field. Firstly, we assume that the vector field is ``light", i.e. $\tilde{m} t \rightarrow 0$. Hence the solution (\ref{Bessel Sol}) is approximated by
\begin{equation} \label{Light B}
B (t) \approx \tilde{c}_1 a^{3 (1 + w) (u + v) / 2} + \tilde{c}_2 a^{3 (1 + w) (u - v) / 2} \,,
\end{equation}
where $\tilde{c}_1$ and $\tilde{c}_2$ are constants. The latter solution means that the evolution of the light vector field is described by a power law for the scale factor. On the other hand, for a ``heavy'' vector field we take the limit of Eq. (\ref{Bessel Sol}) when $\tilde{m} t \rightarrow \infty$ obtaining
\begin{equation}
B(t) \approx a^{ - 3 / 2} \left[ b_1 \, \text{cos} (\tilde{m} t - \varphi) + b_2 \, \text{sin} (\tilde{m} t - \varphi) \right] \,,
\end{equation}
where $b_1$ and $b_2$ are constants. This solution shows that a heavy vector field oscillates with phase $\varphi$ (which is a function of $v$) and envelope decreasing as
\begin{equation} \label{Average B}
\overbar{ B (t) } \propto a^{-3/2} \,.
\end{equation}
This shows that the vector field performs rapid oscillations, hence its dynamical behavior is effectively characterized by the envelope.

\section{Evolution of the energy density}\label{energy}

In the last section, we showed that the vector field has a constant magnitude during inflation. After inflation, it follows either a power law or an oscillatory motion depending whether it is light or heavy, respectively. However, if the vector field is to have a chance to imprint its perturbation spectrum at late times,  it must nearly dominate the universe after inflation, according to the curvaton scenario \cite{Lyth:2001nq}. Therefore, it is necessary to follow the evolution of its energy density.

Varying the corresponding action for the Lagrangian in Eq. $(2)$ with respect to the metric $g_{\mu \nu}$, it follows that \cite{Carter05fu,Nojiri:2005vv}:
\begin{equation} \label{Variation g}
\frac{\delta{\cal L}_{\text{EH}}}{\delta g_{\mu\nu}} + \frac{\delta{\cal L}_\text{inf}}{\delta g_{\mu\nu}} + \frac{\delta{\cal L}_A}{\delta g_{\mu\nu}} + \frac{\delta{\cal L}_{\cal G}}{\delta g_{\mu\nu}} = 0 \,,
\end{equation}
where
\begin{align}
\frac{\delta{\cal L}_{\text{EH}}}{\delta g_{\mu\nu}} & = - \, \frac{m_\text{P}^2}2\left(-R^{\mu\nu}+\frac{1}{2}g^{\mu\nu}R\right) \,,\\
\frac{\delta{\cal L}_A}{\delta g_{\mu\nu}} &= -\frac18 g^{\mu\nu} F_{\alpha\beta} F^{\alpha\beta} + \frac12 F^{\mu\rho}F^\nu_{\ \ \rho} \nonumber \\
 &- \frac12 \, \tilde m^2 \left( A^\mu A^\nu - \frac12 \, g^{\mu\nu} A^\sigma A_\sigma \right) \,,\\
\frac{\delta{\cal L}_{\cal G}}{\delta g_{\mu\nu}} &= - \frac12 \, \alpha \, {\cal G} A^\mu A^\nu + \frac12 \, g^{\mu\nu} f {\cal G} - 2 f R R^{\mu\nu} \nonumber \\
&+ 2 \nabla^\mu \nabla^\nu (f R) - 2 g^{\mu\nu}
\Box(f R) + 8 f R^\mu_{\ \ \rho} R^{\nu\rho} \nonumber\\
&- 4 \nabla_\rho \nabla^\mu (f R^{\nu\rho})
- 4 \nabla_\rho \nabla^\nu (f R^{\mu\rho}) + 4 \Box(f R^{\mu\nu}) \nonumber \\
&+ 4 g^{\mu\nu} \nabla_\rho \nabla_\sigma (f R^{\rho\sigma}) - 2 f R^{\mu\rho\sigma\tau} R^\nu_{\ \ \rho\sigma\tau} \nonumber \\
&+ 4 \nabla_\rho \nabla_\sigma (f R^{\mu\rho\sigma\nu})\,,
\end{align}
and \mbox{$f\equiv\frac12\,\alpha A_\mu A^\mu$}. Using the FLRW metric, the $``00"$  component of Eq. (\ref{Variation g}) can be written as $3 \, m_\text{P}^2 \, H^2 = \rho_{\rm inf} + \rho_B$, where $\rho_{\rm inf}$ is the energy density of the source driving inflation and
\begin{align} 
\rho_B &= \frac12 \dot B^2 + \frac12\left[\tilde m^2 + H^2 \left(1 + 2\frac{\dot B}{B H}\right) \right] B^2\nonumber\\
&+24\alpha H^4 B^2 \left( \frac{\dot B}{H B} + \frac{\dot\alpha}{2 \alpha H} \right) \label{Density B} \,,
\end{align}
is the energy density of the physical vector field $B_\mu = (0, 0, 0, B)$. This is to be compared with the energy density $\rho_B = \frac12 \dot B^2 + \frac12 \,\tilde{m}^2 B^2$ of a vector field non-minimally coupled to gravity through the Ricci scalar \cite{Dimopoulos08a, Karciauskas10as, Golovnev08, Golovnev09a, Golovnev09b}. In our case, the energy density of the vector field has an extra term coming from the coupling with the Gauss-Bonnet invariant.

During inflation, Eq. (\ref{Density B}) gives
\begin{equation}
\rho_B \simeq \frac{1}{2} H^2 B^2\simeq  {\rm const.}\,, 
\end{equation}
where we used the fact that $\tilde{m} \ll H$ and $B$, $H$ and $\alpha$ are nearly constants. We can see that the energy density of the vector field is not diluted by inflation since it remains almost constant.

In order to avoid anisotropic expansion after inflation, the contribution to the energy tensor coming from the vector field must not introduce anisotropic pressures. This can be achieved if the vector field condensate behaves as a pressureless matter, i.e. $\rho_B \propto a^{-3}$  \cite{Dimopoulos06}. In this section we investigate the available parameter space for the constant $\gamma$ that allows this behavior in a radiation dominated universe characterized by $w \approx 1 / 3$. Before to continue our discussion, we want to point out the following. During inflation, the Gauss-Bonnet term is positive and it can be approximated by $\mathcal{G} \approx 24 H^4$.  On the other hand, in order to get a nearly flat power spectrum for the transverse modes we have $\alpha H^2 \approx - 1 / 12$, and therefore $\alpha < 0$. After inflation, when the Hubble parameter $H(t)$ is given by Eq. (\ref{Hubble}), the Gauss-Bonnet invariant is given by
\begin{equation}
{\cal G} =-\frac{64}{27}\frac{1+3w}{t^4(1+w)}.
\end{equation}
So, the Gauss-Bonnet invariant  is negative if $w > - 1 / 3$ (eg. a matter ($w=0$) or a radiation $(w=1/3)$ fluid).  This implies  $\gamma < 0$ and a change in the sign of the effective mass, $m^2 = \tilde{m}^2 + \alpha \mathcal{G}$, from negative, during inflation, to positive, after inflation. As explained in \cite{Dimopoulos06}, a minimally-coupled vector plays the role of a curvaton if it has a negative mass-squared (explicitly $m^2 \approx - 2 H^2$) during inflation. After inflation, the mass-squared has to become positive so that the vector field engages into oscillations and thus avoiding the production of large background anisotropy. As we show, in our model, this change of sign is tacitly provided by the ``evolution'' of the Gauss-Bonnet invariant.

Regarding a light field, we showed in Sec. \ref{evoaf} that the dynamics of the vector field is described as a power law in the scale factor (see Eq. (\ref{Light B})). The power is given in terms of the equation of state parameter  $w$ and the constant $\gamma$. Then, replacing Eqs. (\ref{Hubble}) and (\ref{Light B}) in Eq. (\ref{Density B}), and considering that the dominant fluid after inflation can be either a stiff fluid ($w = 1$) or a radiation fluid ($w = 1 / 3$), one can realize that is impossible to satisfy the condition that  $\gamma < 0$ and the condition that the density of the vector field scales as pressureless matter, so we discard the light field solution.

For a heavy field, we showed that it is oscillating with decreasing envelope as $a^{-3 / 2}$. Following Ref. \cite{Dimopoulos06}, the period of oscillations is much smaller than the Hubble time, which means that the effective behaviour of the vector field is given by its envelope. Therefore, replacing Eq. (\ref{Average B}) in the density in Eq. (\ref{Density B}) we get the average density over many oscillations as
\begin{equation}\label{energyaf}
\overbar{\rho_B} = B_0^2 \left( \frac{1}{32} + 3 \gamma \right)a^{-7} + \frac{1}{2} \tilde{m}^2 B_0^2 a^{-3} \,,
\end{equation}
where $B_0$ is the constant of proportionality implicit in Eq. (\ref{Average B}). The first term in Eq. (\ref{energyaf}) goes as $a^{-7}$, so it decays faster than the radiation dominant fluid (which decays as $a^{-4}$) and even faster than the second term. This means that $\overbar{\rho_B} \propto a^{-3} $. The fact that the average energy density decays as $a^{-3}$ implies two important things: $i)$ the vector field may eventually dominate the Universe and imprint its perturbation spectrum, and $ii)$ the average energy density of the vector field scales as pressureless matter, so the average pressure is zero and therefore there is no generation of large background anisotropy. 

\section{Conclusions}\label{sec-e}

In this paper, we have examined the evolution of a cosmological vector field coupled to the Gauss-Bonnet invariant. Assuming that $\tilde{m} \ll H$, we found that, in order to get a nearly flat power spectrum for the transverse modes during inflation, the coupling $\alpha$ must satisfy the condition $\alpha H^2 \approx - 1/12$. This implies that the power spectrum for the longitudinal modes is $\mathcal{P}_\parallel = 2 \mathcal{P}_{\text{ L, R}}$. Consequently, we showed that the amount of statistical anisotropy in our model, quantified by the parameter $g$, is within the observational bounds, given that the vector field is subdominant during inflation. We also found that the vector field remains constant during the inflationary phase, but it performs rapid oscillations after that whenever $\tilde{m} \gg H$. Averaging over many oscillations, the vector field effectively decays as $a^{- 3 / 2}$, which means that the average energy density, $\overbar{\rho_B}$,  decays as $a^{- 3}$, so, after inflation, the vector field behaves like a pressureless fluid. This indicates that the vector field has a chance to nearly dominate the universe after inflation, without introducing large background  anisotropy, and thus be able to imprint its curvature perturbation.

According to the vector curvaton scenario, the mass of the vector must be negative during inflation and positive after that. In our model, this feature is provided by the particular behavior of the Gauss-Bonnet invariant because it changes its sign when passing from inflation to a matter or radiation dominated epoch. Therefore, we conclude that this non-minimal coupling between a vector field and the Gauss-Bonnet invariant can be a reliable and realistic vector curvaton model.

\section*{Acknowledgments}
This work was supported by COLCIENCIAS Grant 110671250405 and Vi\-ce\-rrec\-tor\'ia de Investigaciones (Univalle) grant  71220. CAV-T thanks Yeinzon Rodr\'iguez for directing our attention to this work.

\appendix

\section{Equations of motion for the transverse and longitudinal modes} \label{Appendix}

In this appendix we outline some details of the steps required to compute the power spectrum of the transverse and longitudinal modes. We start with the background equations of motion. 

Taking $\nu = 0$ in Eq. (\ref{EoM vector}), we get
\begin{equation} \label{Temporal component}
\partial^i \dot{A}_i - \partial^k \partial_k A_t + (a m)^2 A_t = 0 \,,
\end{equation}
and taking $\nu = i$ 
\begin{align} \label{Spatial component}
\ddot{A}^i + H \dot{A}^i + m^2 A^i &= a^{-2}\left[ \partial_k \partial^k A^i - \partial^i \partial^k A_k \right]  \nonumber \\
 &+ \partial^i \left( \dot{A}_t + H A_t \right) \,.
\end{align}
Contracting Eq. (\ref{EoM vector}) with $\partial_\mu$, we obtain an integrability condition which reads
\begin{equation}
(a m)^2 \dot{A}_t - m^2 \partial^i A_i + 3 H \left( \partial^k \partial_k A_t - \partial^i \dot{A}_i \right) = 0 \,.
\end{equation} 
Combining the above equation with Eq. (\ref{Temporal component}), and replacing in Eq. (\ref{Spatial component}), we obtain
\begin{equation} \label{EoM combined}
\ddot{A}^i + H \dot{A}^i + m^2 A^i - a^{-2} \partial^k \partial_k A^i = - 2 H \partial^i A_t \,.
\end{equation}
Since inflation homogenizes the vector field, $\partial_i A_\mu = 0$, hence Eq. (\ref{Temporal component}) implies $A_t = 0$, while the spatial components obey Eq. (\ref{EoM A}).

Now, we perturb around the homogeneous components as in Eq. (\ref{Perturbative expansion}). At first order, the perturbations $\delta A_\mu$ obey the same equations of motion given in Eq. (\ref{Temporal component}) and (\ref{EoM combined}) since they are linear.
Let us switch to Fourier space by expanding the perturbations as
\begin{equation}
\delta A_\mu (t, \boldsymbol{x}) \equiv \int \frac{\text{d}^3 k}{(2 \pi)^{3 / 2}} \delta \mathcal{A}_\mu (t, \boldsymbol{k}) e^{i \boldsymbol{k} \cdot \boldsymbol{x}} \,.
\end{equation}
Inserting this in Eq. (\ref{Temporal component}) we get the following constraint
\begin{equation}
\delta \mathcal{A}_t + i \frac{k^j \delta \mathcal{\dot{A}}_j}{k^2 + (a m)^2} = 0 \,,
\end{equation}
which allows us to write $\delta \mathcal{A}_t$ in terms of $\delta \mathcal{A}_i$. Using this in Eq. (\ref{EoM combined}) we get
\begin{equation}
\delta \ddot{\mathcal{A}}^i+ H \delta \dot{\mathcal{A}}^i + \left[ m^2 + \left( \frac{k}{a} \right)^2 \right] \delta \mathcal{A}^i + 2 H k^i \frac{k^j \delta \dot{\mathcal{A}}_j}{k^2 + (a m)^2} = 0 \,. 
\end{equation}
The last step consists in defining the longitudinal and transverse components as
\begin{equation}
\delta \mathcal{A}^i_\parallel \equiv k^i \left( \frac{k^j \delta \mathcal{A}_j}{k^2} \right) \,, \ \delta \mathcal{A}^i_\perp \equiv \delta \mathcal{A}^i - \delta \mathcal{A}^i_\parallel \,,
\end{equation}
and replacing in the latter equation, we get Eqs. (\ref{Transverse A}) and (\ref{Longitudinal A}).

\bibliography{BiblioGB}

\end{document}